\documentclass[12pt]{article}
\usepackage{psfrag,amsmath,epsfig,feynarts,a4,cite,latexsym}
\def\pslash#1{{\setbox0=\hbox{$#1$}
  \rlap{\ifdim\wd0>.7em\kern.22\wd0\else\kern.1\wd0\fi /}#1}}
\def\psl{\pslash p}
\def\ksl{\pslash k}
\def\qsl{\pslash q}

\def\glui{{\tilde{g}}}
\def\epsbar{{\bar{\epsilon}}}

\allowdisplaybreaks[1]
\begin{document}
\begin{flushright}
IPPP/05/53\\
DCPT/05/106\\
MPP--2005--108\\
\end{flushright}
\vspace{3em}
\begin{center}
{\large\bf MSSM Higgs-boson mass predictions and
}\\[2ex]
{\large\bf 
two-loop non-supersymmetric counterterms  
}
\\
\vspace{3em}
{\sc Wolfgang Hollik$^1$, Dominik St\"ockinger$^2$
}\\[2em]
{\it ${}^1$\ Max-Planck-Institut f\"ur Physik
             (Werner-Heisenberg-Institut) \\
              D-80805 Munich, Germany        \\[1em]
${}^2$\ Institute for Particle Physics Phenomenology, Physics,
\\ University of Durham,
Durham DH1~3LE, UK
}
\setcounter{footnote}{0}
\end{center}
\vspace{2ex}
\begin{abstract}

The evaluation of Yukawa-enhanced two-loop contributions to the MSSM
Higgs-boson mass is considered. We prove the common assumption that 
regularization by dimensional reduction preserves supersymmetry at the
required level. Thus generating counterterms by multiplicative
renormalization is correct. Technically, we identify a suitable
Slavnov-Taylor identity, use a recently developed method to evaluate
it at the two-loop level, and show that it is valid in dimensional
reduction.

\end{abstract}

\section{Introduction}

The prediction of the mass of the lightest Higgs boson is one of the
most striking features of the Minimal Supersymmetric Standard Model
(MSSM). At tree level, the mass of the lightest MSSM Higgs boson $M_h$
is bound to be smaller than the mass of the $Z$-boson. At
higher orders, $M_h$ is a function of all MSSM parameters, but there
still is a very restrictive upper bound as long as the masses of
supersymmetric particles are not much higher than 1~TeV.

The LEP exclusion bound of $114.4$~GeV \cite{LEPBound} for the mass of 
a standard model (SM)-like Higgs boson allows to derive stringent
constraints on the MSSM parameter space. In the future, the direct
measurement of the Higgs-boson mass will lead to very small
experimental uncertainties of only $200$~MeV at the LHC 
\cite{LHCUncertainty} and  $50$~MeV at the ILC
\cite{ILCUncertainty}. Comparing the measurement of $M_h$ to its
predicted value amounts to sensitive tests of the MSSM and allows the
indirect determination of further MSSM parameters. It is therefore 
important to know $M_h$ as a function of the MSSM parameters as
precisely as possible.

A lot of effort has been put into the evaluation of $M_h$ (see Ref.\
\cite{Review} for a review). So far, the one-loop contributions are
completely known. The most important two-loop corrections are the ones
enhanced by the Yukawa couplings of the top and bottom quark
$\alpha_{t,b}=y_{t,b}^2/(4\pi)$. The Yukawa-enhanced contributions of
${\cal O}(\alpha_t\alpha_s)$~\cite{mhiggsEP1b,mhiggsRG1,mhiggsRG2,mhiggsletter,mhiggslong,mhiggsEP0,mhiggsEP1,bse,reconc},
${\cal O}(\alpha_t^2)$~\cite{mhiggsEP1b,mhiggsEP3,mhiggsEP2},
${\cal O}(\alpha_b\alpha_s)$~\cite{mhiggsEP4,heidi},
${\cal O}(\alpha_t\alpha_b)$, ${\cal O}(\alpha_b^2)$~\cite{mhiggsEP5}
are known in the on-shell renormalization scheme. All
these Yukawa-enhanced contributions can be evaluated in the gauge-less
limit, where the electroweak gauge couplings and $M_{W,Z}$ go to zero
with fixed ratio $M_W/M_Z$ and fixed Higgs vacuum expectation
values. Further electroweak two-loop contributions that go beyond the
gauge-less limit are known in the $\overline{DR}$ scheme
\cite{effpotfull,mhiggsp2}. The remaining intrinsic error of the
Yukawa-enhanced 
corrections has been estimated to $\delta M_h^{\rm intr}=3$~GeV
\cite{intrerror}.

There might be, however, an additional Yukawa-enhanced contribution 
from non-supersymmetric counterterms at the two-loop level.
So far, such terms  have been assumed 
to vanish, but without an explicit investigation.
In this paper we close this gap by    
a detailed evaluation of the counterterm structure
at two-loop order.

In all 
the references listed above,
regularization by dimensional reduction (DRED)~\cite{Siegel79} has
been employed and it has been assumed that all symmetries of the
MSSM are preserved by this method.
According to this assumption, the structure of the counterterms is
symmetric and generated by multiplicative renormalization. 
However, although dimensional reduction has been checked to preserve
supersymmetry in many individual cases, there is no general proof. 

The traditional checks apply to one-loop self energies
\cite{CJN80}, one-loop on-shell \cite{BHZ96} and off-shell
\cite{STIChecks1,STIChecks2} three-point functions and further
one-loop relations involving soft and spontaneous symmetry
breaking \cite{STIChecks3}. Further checks concern the renormalization
group $\beta$ functions at the two-loop level \cite{betachecks},
corresponding to divergences of two-loop diagrams.  
Most recently,
supersymmetry identities for one-loop four-point functions and
two-loop two-point functions (without spontaneous symmetry breaking)
have been checked \cite{DREDPaper}. In all cases, dimensional
reduction was in agreement with the required supersymmetry relations. 
These checks are indeed sufficient to prove that multiplicative
renormalization is correct for the one-loop counterterms of the Higgs,
quark and squark sectors entering the two-loop calculation of $M_h$
at the level of one-loop subrenormalization.

These results, however, do not constitute a proof that dimensional
reduction preserves supersymmetry at the two-loop level. In
particular, they cannot exclude that 
supersymmetry is broken by a finite amount, 
which could be relevant for precision calculations of $M_h$.

In the case of supersymmetry-breaking by the method of regularization,
extra symmetry-restoring counterterms  have to be added as
discussed e.g.\ in Refs.\
\cite{STIChecks1,STIChecks2,STIChecks3}. These extra counterterms,
which are by themselves  
not supersymmetric, are required to restore the original symmetry
at the quantum level at the considered order.
They are finite since all divergences
are cancelled by symmetric counterterms derived from multiplicative
renormalization.
If needed, they would cause a finite shift in the Higgs-boson 
self-energies and thus
would be relevant for precision calculations of $M_h$.
It is therefore important to clarify the necessity 
for such counterterms.

In particular, if the symmetry is broken, the two-loop counterterms
for the Higgs-potential 
have to be
modified according to 
\begin{align}
\delta^{(2)} V &= \delta^{(2)} V_{\rm sym}+
\delta^{(2)} V_{\rm non-susy},
\label{dVgeneral}
\end{align}
which contains the usual multiplicative symmetric counterterms
$\delta^{(2)}V_{\rm sym}$ and additional supersymmetry-restoring
counterterms $\delta^{(2)}V_{\rm non-susy}$. 
The renormalizability of the MSSM \cite{SSTI1,SSTI2,SSTIus1,SSTIus2}
guarantees that supersymmetry can be restored by a suitable choice of
$\delta^{(2)}V_{\rm non-susy}$. But since the structure of these extra
counterterms does not correspond to multiplicative
renormalization, they have to be derived from the supersymmetric
Slavnov-Taylor identities explicitly evaluated at the two-loop level.

In the present paper we derive the relevant supersymmetric 
two-loop Slavnov-Taylor identities that are potentially affected by
$\delta^{(2)} V_{\rm non-susy}$ and determine the values
of the
supersymmetry-restoring counterterms by a direct calculation.
Since we are aiming at contributions from the large 
Yukawa couplings, we restrict ourselves to
the gauge-less limit with respect to the electroweak interaction,
i.e.\ we take into account electroweak Yukawa interactions and
the strong interaction in terms of supersymmetric QCD.

Our explicit determination makes use of recent developments
concerning dimensional reduction \cite{DREDPaper,DREDProc} (see also
Refs.\ \cite{Harlander05,Fact} for complementary recent work). In Ref.\
\cite{DREDPaper}, it has been shown that a slight reformulation of
dimensional reduction, which does not affect its application to the
calculation of $M_h$, avoids the mathematical inconsistency of Ref.\
\cite{Siegel80}. Based on this reformulation, a dramatic
simplification for evaluating supersymmetric Slavnov-Taylor
identities has been obtained. The evaluation of such
identities constitutes the central part of our calculation of
$\delta^{(2)}V_{\rm non-susy}$.

The general strategy of our calculation follows the approach described
in Refs.\ \cite{STIChecks1,STIChecks2,STIChecks3}. 
First we discuss
general properties of the potential supersymmetry-restoring 
counterterms that are relevant for the calculation of the Higgs-boson
mass. Then we derive suitable Slavnov-Taylor identities that
are sensitive to these counterterms. Finally, 
the Slavnov-Taylor identities are evaluated
using the method of Ref.~\cite{DREDPaper}.

\section{Higgs potential and counterterms}

The two-loop evaluation of the Higgs-boson mass spectrum
requires three types of counterterms: 
one-loop counterterms that renormalize divergent
one-loop subdiagrams, products of one-loop counterterms that partially
renormalize two-loop diagrams, and genuine two-loop counterterms. 
As mentioned in the introduction, multiplicative renormalization is
sufficient for the one-loop counterterms. Here we 
examine whether this applies also to the genuine two-loop
counterterms.

The genuine two-loop counterterms relevant for the evaluation of the
Higgs-boson self-energies
correspond to the MSSM Higgs potential $V$. At tree
level, $V$ is given by
\begin{subequations}
\begin{align}
V &= V_{\rm quadratic}+V_{\rm quartic},\\
\label{Vquad}
V_{\rm quadratic} &=
{m}_1^2 |{\cal H}_1|^2 + {m}_2^2 |{\cal H}_2|^2
    + m_3^2 
\left({\cal H}_1^1 {\cal H}_2^2-{\cal H}_1^2 {\cal H}_2^1
 + h.c.\right),\\
V_{\rm quartic} &=\frac{g_1^2+g_2^2}{8}
\left(\left|{\cal H}_1\right|^2-\left|{\cal H}_2\right|^2\right)^2
 + \frac{g_2^2}{2}\left|{\cal H}_1^\dagger {\cal H}_2\right|^2 
\end{align}
\end{subequations}
in terms of the Higgs doublets ${\cal H}_1=H_1+{v_1\choose0}$, 
${\cal H}_2=H_2+{0\choose v_2}$, their respective vacuum
expectation values $v_{1,2}$, and the $SU(2)$ and $U(1)$ gauge
couplings $g_{2,1}$. The mass parameters $m_{1,2}^2$ are combinations
of soft supersymmetry-breaking parameters and the $\mu$-parameter, and
$m_3^2$ is a soft supersymmetry-breaking parameter.

The symmetric counterterms used in the literature are generated by the
multiplicative renormalization transformation
\begin{subequations}
\begin{align}
m_i^2 &\to m_i^2+\delta^{(1)} m_i^2+\delta^{(2)} m_i^2,\\
g_i   &\to g_i +\delta^{(1)} g_i+\delta^{(2)} g_i,\\
v_i   &\to v_i +\delta^{(1)} v_i+\delta^{(2)} v_i,\\
H_i   &\to (1 + \delta^{(1)} Z_{H_i}+\delta^{(2)} Z_{H_i})^{1/2}
           H_i,
\end{align}
\end{subequations}
where the notation $\delta^{(l)}$ refers to a renormalization constant of
$l$-loop order. Thus, the genuine two-loop symmetric counterterms have the
structure
\begin{align}
\delta^{(2)} V_{\rm sym} &= \left[\delta^{(2)}
m_i^2\partial_{m_i^2} 
+\delta^{(2)} g_i\partial_{g_i}
+\frac12 H_i\delta^{(2)} Z_{H_i}\partial_{H_i}+\delta^{(2)}
v_i\partial_{v_i}\right]
V
,
\label{dVmult}
\end{align}
which is sufficient only if DRED preserves
supersymmetry at the two-loop level in the gauge-less limit.

If this is not the case, $\delta^{(2)}V$ has to be augmented by
supersymmetry-restoring counterterms, as indicated in~(\ref{dVgeneral}). 
A priori, $\delta^{(2)}V_{\rm non-susy}$ contains
quadratic and quartic terms,
\begin{align}
\delta^{(2)}V_{\rm non-susy}&=
\delta^{(2)}V_{\rm non-susy}^{\rm quadratic}+
\delta^{(2)}V_{\rm non-susy}^{\rm quartic}.
\end{align}
Two simple properties of $\delta^{(2)} V_{\rm non-susy}$ can 
immediately be derived. First, $\delta^{(2)} V_{\rm non-susy}$ respects
global $SU(2)\times U(1)$ gauge invariance. The reason is that
DRED\footnote{%
The version of DRED used in the literature is the one defined in 
Ref.~\cite{DREDPaper} with anticommuting $\gamma_5$.}
preserves global gauge invariance as can easily be obtained e.g.\ from
the gauge invariance of the regularized Lagrangian using the quantum
action principle~\cite{DREDPaper}. As a consequence,
$\delta^{(2)}V_{\rm non-susy}$ depends on the Higgs doublets $H_{1,2}$
and $v_{1,2}$ only via the combinations
${\cal H}_{1,2}$,
similar to $V$ itself. 
Second, the quadratic terms in $\delta^{(2)}
V_{\rm non-susy}$ are not necessary since the quadratic terms in the
Higgs potential (\ref{Vquad}) are not restricted by supersymmetry in
the first place, and no supersymmetry-breaking could modify their
structure.
Therefore, we have 
\begin{align}
\delta^{(2)}V_{\rm non-susy}^{\rm quadratic} &=0.
\label{dVquad}
\end{align}

\section{Relevant Slavnov-Taylor identities}

Any non-zero counterterm quartic in ${\cal H}_1, {\cal H}_2$
modifies both
the Higgs-boson four-point and two-point vertex functions and can thus
contribute to the Higgs-boson mass prediction. For example, the term 
$|{\cal H}_1|^4=
|H_1|^4+2v_1^2|H_1|^2+4v_1^2({\rm Re}H_1^1)^2+\ldots$ 
would contribute to the four-point function $\Gamma_{H_1 H_1^\dagger H_1 H_1^\dagger}$
and to the two-point functions involving $H_1$.
On the other hand, owing to (\ref{dVquad}), vanishing non-supersymmetric
quartic counterterms would imply also the 
absence of non-supersymmetric counterterms for the self-energies.
It is therefore sufficient to evaluate
the quartic non-supersymmetric counterterms 
$\delta^{(2)}V_{\rm non-susy}^{\rm quartic}$ 
by considering their contribution to four-point functions.

The starting point for the evaluation is the basic requirement that
{\em after} renormalization, i.e.\ in particular after adding
$\delta^{(2)} V^{\rm quartic}_{\rm non-susy}$, the 
supersymmetric Slavnov-Taylor identity of the MSSM~\cite{SSTIus2} holds,
\begin{align}
S(\Gamma)&=0.
\end{align}
Here $\Gamma$ denotes the renormalized effective action, the
generating functional of the renormalized one-particle irreducible
vertex functions. 

The following derivative of the Slavnov-Taylor identity directly
constrains the desired Higgs-boson four-point functions:
\begin{align}
0&=\frac{\delta^5  S(\Gamma)}{\delta\phi_a\delta\phi_b\delta\phi_c
\delta\tilde{H}_{kL}^l\delta\bar\epsilon_L},
\label{phi4STI1}
\end{align}
where $\phi_i$ denote any components of the MSSM Higgs bosons $H_i^j$,
$H_i^j{}^\dagger$, and $\tilde{H}_k^l$ is the Higgsino partner of
$H_k^l$. $\epsilon$ denotes the supersymmetry ghost. The index $L$
denotes left-handed spinors,
\begin{align}
\frac{\delta}{\delta \bar\epsilon_L}&=
P_L\frac{\delta}{\delta \bar\epsilon},&
\frac{\delta}{\delta \tilde{H}_{kL}^l}&=
\frac{\delta}{\delta \tilde{H}_k^l{}}P_L
\end{align}
with $P_L=\frac12(1-\gamma_5)$. In the gauge-less limit, identity
(\ref{phi4STI1}) is equivalent to the following relation between
renormalized vertex functions, 
\begin{align}
0=
\sum_{\phi_i}\bigg[
&
\Gamma_{\tilde{H}_{kL}^lY_{\phi_i}\bar\epsilon_L}
\Gamma_{\phi_a\phi_b\phi_c\phi_i}
+
\Gamma_{\phi_a\phi_b\phi_c\tilde{H}_{kL}^lY_{\phi_i}\bar\epsilon_L}
\Gamma_{\phi_i}
\nonumber\\
+\bigg(
&
\Gamma_{\phi_a\tilde{H}_{kL}^lY_{\phi_i}\bar\epsilon_L}
\Gamma_{\phi_b\phi_c\phi_i}
+
\Gamma_{\phi_a\phi_b\tilde{H}_{kL}^lY_{\phi_i}\bar\epsilon_L}
\Gamma_{\phi_c\phi_i}
+\mbox{perm}\bigg)\bigg]
\nonumber\\
-
\sum_{i,j}\bigg[
&
\Gamma_{y_i^j\bar\epsilon_L}
\Gamma_{\phi_a\phi_b\phi_c \tilde{H}_{kL}^l\overline{\tilde{H}}_i^j}
+
\Gamma_{y_i^{jC}\bar\epsilon_L}
\Gamma_{\phi_a\phi_b\phi_c \tilde{H}_{kL}^l\overline{\tilde{H}}_i^{jC}}
\nonumber\\
+
\bigg(&
\Gamma_{\phi_a y_i^j\bar\epsilon_L}
\Gamma_{\phi_b\phi_c \tilde{H}_{kL}^l\overline{\tilde{H}}_i^j}
+
\Gamma_{\phi_a y_i^{jC}\bar\epsilon_L}
\Gamma_{\phi_b\phi_c\tilde{H}_{kL}^l\overline{\tilde{H}}_i^{jC}}
+\mbox{perm}\bigg)\nonumber\\
+
\bigg(&
\Gamma_{\phi_a\phi_b y_i^j\bar\epsilon_L}
\Gamma_{\phi_c\tilde{H}_{kL}^l\overline{\tilde{H}}_i^j}
+
\Gamma_{\phi_a\phi_b y_i^{jC}\bar\epsilon_L}
\Gamma_{\phi_c\tilde{H}_{kL}^l\overline{\tilde{H}}_i^{jC}}
+\mbox{perm}\bigg)\nonumber\\
+
&
\Gamma_{\phi_a\phi_b\phi_c y_i^j\bar\epsilon_L}
\Gamma_{\tilde{H}_{kL}^l\overline{\tilde{H}}_i^j}
+
\Gamma_{\phi_a\phi_b\phi_c y_i^{jC}\bar\epsilon_L}
\Gamma_{\tilde{H}_{kL}^l\overline{\tilde{H}}_i^{jC}}
\bigg]
\nonumber\\
+&\sqrt2 P_L f_0 \Gamma_{\phi_a\phi_b\phi_c \tilde{H}_{kL}^l
\overline{\chi}_L^C}
.
\label{phi4STI}
\end{align}
Although this identity is quite involved, it is easy to see that it
relates the desired quantity, the $\phi^4$-interaction, which appears
as the first term in the first line, to many other vertex functions. In
the first sum, $\phi_i$  runs over all Higgs boson components $H_i^j$,
$H_i^j{}^\dagger$, and in the second sum  $i,j$ take the values
$1,2$. The abbreviation ``perm'' denotes terms 
corresponding to all possible permutations of $\phi_{a,b,c}$. The
conventions for the Slavnov-Taylor identity of the MSSM have been
adapted to four-spinors as e.g.\ in Refs.\
\cite{STIChecks2,STIChecks3}. The symbols $Y_{\phi_i}$ and $\bar{y}_i^j$
denote the sources of the BRS variations of $\phi_i$ and
$\tilde{H}_i^j$, respectively; $\chi$ and $f_0$ are the spinorial and
constant components of the spurion superfield introduced to describe
soft supersymmetry breaking \cite{SSTIus1}.

At the two-loop level, each term of the form $\Gamma_A\Gamma_B$ can be
decomposed as follows,
\begin{align}
\Gamma^{(0)}_A
\left(\Gamma^{(2),\rm DRED}_B+\Gamma^{(2),\rm ct}_B\right)+
\Gamma^{(1)}_A\Gamma_B^{(1)}+
\left(\Gamma^{(2),\rm DRED}_A+\Gamma^{(2),\rm ct}_A\right)
\Gamma_B^{(0)},
\end{align}
where $\Gamma^{(l)}$ denotes the $l$-loop order contribution. Thereby,
the two-loop contribution has been split into
the regularized vertex functions including one-loop but excluding
two-loop counterterms (denoted by the upper index ``${\rm DRED}$''), and
the 
corresponding genuine two-loop counterterms (index ``${\rm ct}$''). 

Likewise, identity (\ref{phi4STI1}) or (\ref{phi4STI}), evaluated
at $2$-loop order, can be rewritten in the form
\begin{align}
0 &= \Delta^{(2),\rm ct}+\Delta^{(2),\rm DRED}
,
\label{phi4STI3}
\end{align}
where $\Delta^{(2),\rm ct}$ is the contribution from genuine two-loop
counterterms and $\Delta^{(2),\rm DRED}$ is the remaining contribution,
involving the regularized vertex functions and one-loop counterterms.
In the gauge-less limit, the first contribution in~(\ref{phi4STI3}) 
is given by
\begin{align}
\Delta^{(2),\rm ct}=
\sum_{\phi_i}
\bigg[
&
\Gamma^{(0)}_{\tilde{H}_{kL}^lY_{\phi_i}\bar\epsilon_L}
\Gamma^{(2),\rm ct}_{\phi_a\phi_b\phi_c\phi_i}
+\Gamma^{(2),\rm ct}_{\tilde{H}_{kL}^lY_{\phi_i}\bar\epsilon_L}
\Gamma^{(0)}_{\phi_a\phi_b\phi_c\phi_i}\bigg]
\nonumber\\
+
\sum_{i,j}
\bigg[
&\Gamma^{(2),\rm ct}_{\phi_a\phi_b y_i^j\bar\epsilon_L}
\Gamma^{(0)}_{\phi_c\tilde{H}_{kL}^l\overline{\tilde{H}}_i^j}
+
\Gamma^{(2),\rm ct}_{\phi_a\phi_b y_i^{jC}\bar\epsilon_L}
\Gamma^{(0)}_{\phi_c\tilde{H}_{kL}^l\overline{\tilde{H}}_i^{jC}}
\nonumber\\
+&
\Gamma^{(0)}_{\phi_a\phi_b y_i^j\bar\epsilon_L}
\Gamma^{(2),\rm ct}_{\phi_c\tilde{H}_{kL}^l\overline{\tilde{H}}_i^j}
+
\Gamma^{(0)}_{\phi_a\phi_b y_i^{jC}\bar\epsilon_L}
\Gamma^{(2),\rm ct}_{\phi_c\tilde{H}_{kL}^l\overline{\tilde{H}}_i^{jC}}
+\mbox{perm}\bigg]
.
\label{Deltact}
\end{align}
All terms in~(\ref{phi4STI}) that do not re-appear in~(\ref{Deltact})
vanish
since they involve Green functions of power-counting dimension${}\ge5$,
for which there are neither classical nor counterterm
contributions. Moreover, most of the terms in (\ref{Deltact})
vanish due to global gauge invariance or as a result of the gauge-less
limit. The only term that can be non-zero is the first term for
$\phi_i=\phi_d\equiv H_k^l$. The result for
$\Gamma^{(0)}_{\tilde{H}_{kL}^lY_{\phi_d}\bar\epsilon_L}$ is well
known (see e.g.\ \cite{STIChecks2,STIChecks3}), and we obtain 
\begin{align}
\Delta^{(2),\rm ct} &= \sqrt{2}P_L
\Gamma^{(2),\rm ct}_{\phi_a\phi_b\phi_c\phi_d}\ .
\end{align}

The second contribution in~(\ref{phi4STI3}) has the same form as
the right-hand side of~(\ref{phi4STI}), but the products
$\Gamma_A\Gamma_B$ have to be evaluated at the two-loop level 
without the genuine two-loop counterterms. $\Delta^{(2),\rm DRED}$ can
be written in the following simple form,
\begin{align}
\Delta^{(2),\rm DRED}&=
\left[\frac{\delta^5  S(\Gamma^{(2),\rm DRED})}{\delta\phi_a\delta\phi_b
\delta\phi_c\delta\phi_d\delta\bar\epsilon_L}\right]_{\rm 2-loop},
\label{DeltaDRED}
\end{align}
where $\Gamma^{(2), \rm DRED}$ denotes the regularized vertex functional
including one-loop but excluding $2$-loop counterterms. 

Hence, identity (\ref{phi4STI1}) leads to
\begin{align}
P_L\Gamma^{(2),\rm ct}_{\phi_a\phi_b\phi_c \phi_d}
&=-\frac{1}{\sqrt2}\,
\left[\frac{\delta^5  S(\Gamma^{(2),\rm DRED})}{\delta\phi_a\delta\phi_b
\delta\phi_c\delta\phi_d\delta\bar\epsilon_L}\right]_{\rm 2-loop},
\label{phi4ctId}
\end{align}
at the two-loop level in the gauge-less limit. This is an equation
that directly determines the counterterm for any
$\phi^4$-interaction. Since the multiplicative counterterm to $\Gamma^{\rm
  ct}_{\phi_a\phi_b\phi_c\phi_d}$ vanishes  in the gauge-less limit,
we can rewrite this equation in the form
\begin{align}
\frac{\delta^4}
{\delta\phi_a\delta\phi_b\delta\phi_c\delta \phi_d}
\int d^Dx\,P_L\,\delta^{(2)}V^{\rm quartic}_{\rm non-susy}
&=
\frac{1}{\sqrt2}\,
\left[\frac{\delta^5  S(\Gamma^{(2),\rm DRED})}{\delta\phi_a\delta\phi_b
\delta\phi_c\delta\phi_d\delta\bar\epsilon_L}\right]_{\rm 2-loop}.
\label{phi4ctId2}
\end{align}
Obviously, by choosing all possible combinations of $\phi_i$, this
equation, together with its complex conjugate, completely determines
the non-supersymmetric counterterms at the desired order.

\section{Evaluation of the Slavnov-Taylor identities}
\label{sec:eval}

The remaining task is to evaluate the right-hand side of~(\ref{phi4ctId2}). 
Traditionally, we would have to
evaluate all Green functions appearing in~(\ref{phi4STI}), in
particular, also 5-point functions at the two-loop level. 
We can simplify the evaluation by using the results of
Ref.\ \cite{DREDPaper}, according to which the right-hand side 
of~(\ref{phi4ctId2}) is equal to
\begin{align}
\Delta^{(2),\rm DRED}&= -i P_L
\left[\left([S(\Gamma^{(0)})]\cdot
\Gamma^{\rm DRED}\right)_{\phi_a\phi_b\phi_c\tilde{H}_k^l\bar\epsilon}
\right]_{\rm 2-loop}
P_L 
,
\label{sreg}
\end{align}
involving an insertion of the
operator $[S(\Gamma^{\rm (0)})]$ at the two-loop level in the
gauge-less limit.

In Ref.\ \cite{DREDPaper}, the operator $[S(\Gamma^{(0)})]$ has been
given for a general supersymmetric gauge theory without
spontaneous breaking of gauge invariance and soft supersymmetry
breaking. For the purpose of the present paper we have rederived this
operator for the case of the MSSM, taking into account both symmetry
breakings in the way defined in Ref.\ \cite{SSTIus2}. It turns out
that those parts that can play a role in~(\ref{sreg}) are not
modified compared to the case without symmetry breaking.
All relevant terms in $[S(\Gamma^{(0)})]$ are four-fermion operators
involving $\bar\epsilon$ and three other fermions (here either
gluinos, Higgsinos, or quarks). 

\begin{figure}[tb]
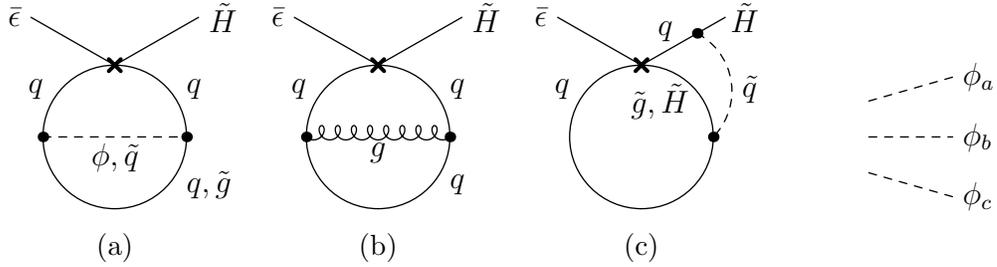

\begin{center}
\unitlength=1.cm%
\begin{feynartspicture}(14,4)(4,1)
\FADiagram{(a)}
\FAVert(10,16){1}
\FALabel(3,20)[r]{$\epsbar$\ }
\FAProp(17,20)(10,16)(0.,){/Straight}{0}
\FALabel(17,20)[l]{\ ${\tilde{H}}$}
\FAProp(3,20)(10,16)(0.,){/Straight}{0}
\FALabel(4,14)[r]{$q$}
\FALabel(16,14)[l]{$q$}
\FALabel(16,6)[l]{$q,\glui$}
\FALabel(10,9.5)[t]{$\phi,\tilde{q}$}
\FAProp(4,10)(16,10)(0.,){/ScalarDash}{0}
\FAProp(10,16)(10,4)(1.,){/Straight}{0}
\FAProp(10,16)(10,4)(-1.,){/Straight}{0}
\FAVert(4,10){0}
\FAVert(16,10){0}
\FADiagram{(b)}
\FAVert(4,10){0}
\FAVert(16,10){0}
\FAVert(10,16){1}
\FALabel(3,20)[r]{$\epsbar$\ }
\FAProp(17,20)(10,16)(0.,){/Straight}{0}
\FALabel(17,20)[l]{\ ${\tilde{H}}$}
\FAProp(3,20)(10,16)(0.,){/Straight}{0}
\FALabel(4,14)[r]{$q$}
\FALabel(16,14)[l]{$q$}
\FALabel(16,6)[l]{$q$}
\FALabel(10,9.5)[t]{$g$}
\FAProp(4,10)(16,10)(0.,){/Cycles}{0}
\FAProp(10,16)(10,4)(1.,){/Straight}{0}
\FAProp(10,16)(10,4)(-1.,){/Straight}{0}
\FADiagram{(c)}
\FAVert(14.66,18.66){0}
\FAVert(16,10){0}
\FAVert(10,16){1}
\FALabel(3,20)[r]{$\epsbar$\ }
\FAProp(17,20)(10,16)(0.,){/Straight}{0}
\FALabel(17,20)[l]{\,$\tilde{H}$}
\FAProp(3,20)(10,16)(0.,){/Straight}{0}
\FALabel(4,14)[r]{$q$}
\FALabel(18,14)[l]{\,$\tilde{q}$}
\FALabel(14,13)[r]{$\glui,\tilde{H}$}
\FALabel(12,18)[b]{${q}$}
\FAProp(10,16)(10,4)(1.,){/Straight}{0}
\FAProp(10,16)(10,4)(-1.,){/Straight}{0}
\FAProp(14.66,18.66)(16,10)(-.5,){/ScalarDash}{0}
\FADiagram{}
\FAProp(7,7)(14,5)(0.,){/ScalarDash}{0}
\FAProp(7,10)(14,10)(0.,){/ScalarDash}{0}
\FAProp(7,13)(14,15)(0.,){/ScalarDash}{0}
\FALabel(14,5)[l]{\ $\phi_c$}
\FALabel(14,10)[l]{\ $\phi_b$}
\FALabel(14,15)[l]{\ $\phi_a$}
\end{feynartspicture}
\end{center}
\vspace{-.5cm}
\caption{Diagrams contributing to~(\ref{sreg}) at the two-loop
  level in the gauge-less limit. The insertion of the operator
  $[S(\Gamma^{(0)})]$ is marked by a cross. Quarks, gluons and Higgs
  bosons are denoted by $q$, $g$, $\phi$; squarks, gluinos and
  Higgsinos are denoted by $\tilde{q}$, $\glui$, and $\tilde{H}$. 
  The lines corresponding to the external Higgs bosons $\phi_{a,b,c}$
  have to be attached in all possible ways.
}
\label{fig:diagrams}
\end{figure}

Fig.~\ref{fig:diagrams} shows the
Feynman diagrams contributing in the gauge-less limit to (\ref{sreg}),
i.e.\ to the vertex function with external
$\phi_{a,b,c}$, $\tilde{H}_k^l$, $\epsbar$ and the insertion of the
composite operator $[S(\Gamma^{(0)})]$. The
insertion of $[S(\Gamma^{(0)})]$ is marked by a cross, and the three
Higgs fields have to be attached in all possible ways. In all
diagrams, the basic fermion-loop topology is ``Topology (c)'' from
Ref.\ \cite{DREDPaper}; i.e.\ the open fermion line involves
$\bar\epsilon$ and quarks or Higgsinos, but no gauginos.

It is useful to denote the chain of $\gamma$-matrices corresponding to
the open fermion line as $A$, and the $\gamma$-chain corresponding to
the closed fermion loop as $B$. Then the Feynman rules for the
diagrams involve terms like the following ones,
\begin{align}
P_L \gamma^\mu B \gamma_\mu P_L A P_L,\quad
P_L A P_L Tr(P_R B),\quad\ldots
\end{align}
We do not need the detailed Feynman rules but only two properties that
can be easily read off from the rules given in Ref.\ \cite{DREDPaper}
combined with the explicit factors $P_L$ appearing in~(\ref{sreg}):
\begin{enumerate}
\item a diagram vanishes if the number of $\gamma^\mu$-matrices in
  $AB$ is odd.
\item a diagram vanishes if the number of $\gamma^\mu$-matrices in $B$
  is smaller than four.
\end{enumerate}
After carrying out the two loop integrals in diagrams of the form
Fig.~\ref{fig:diagrams}a,b, the only covariants that can appear in
the $\gamma$-chain $B$ are $\psl_{a,b,c}$, where $p_{a,b,c}$ are the
incoming momenta of the Higgs fields $\phi_{a,b,c}$. Hence the
$\gamma$-chain $B$ can be simplified to terms involving at most three
factors of $\psl_i$ and no other $\gamma^\mu$-matrices. As a result of
property 2 above, these diagrams all vanish.

\begin{figure}[tb]
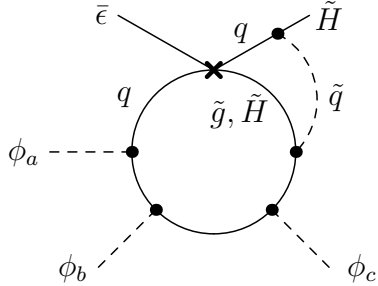

\begin{center}
\unitlength=1.cm%
\begin{feynartspicture}(4,4)(1,1)
\FADiagram{}
\FAVert(14.66,18.66){0}
\FAVert(16,10){0}
\FAVert(10,16){1}
\FALabel(3,20)[r]{$\epsbar$\ }
\FAProp(17,20)(10,16)(0.,){/Straight}{0}
\FALabel(17,20)[l]{\,$\tilde{H}$}
\FAProp(3,20)(10,16)(0.,){/Straight}{0}
\FALabel(4,14)[r]{$q$}
\FALabel(18,14)[l]{\,$\tilde{q}$}
\FALabel(14,13)[r]{$\glui,\tilde{H}$}
\FALabel(12,18)[b]{${q}$}
\FAProp(10,16)(10,4)(1.,){/Straight}{0}
\FAProp(10,16)(10,4)(-1.,){/Straight}{0}
\FAProp(14.66,18.66)(16,10)(-.5,){/ScalarDash}{0}
\FAProp(-2,10)(4,10)(0.,){/ScalarDash}{0}
\FAProp(5.76,5.76)(1.51,1.51)(0.,){/ScalarDash}{0}
\FAProp(14.24,5.76)(18.48,1.51)(0.,){/ScalarDash}{0}
\FALabel(19,1.5)[l]{\ $\phi_c$}
\FALabel(1.5,1.5)[r]{ $\phi_b$\ }
\FALabel(-2,10)[r]{ $\phi_a$\ }
\FAVert(5.76,5.76){0}
\FAVert(14.24,5.76){0}
\FAVert(4,10){0}
\end{feynartspicture}
\end{center}
\vspace{-.5cm}
\caption{Diagram of the form shown in
  Fig.\ \ref{fig:diagrams}c, where all Higgs-boson lines are attached to the
  fermion loop. In this instance the $\gamma$-chains $A$ and $B$ are
  products of 
  one and five fermion propagators, respectively.
}
\label{fig:dangerous}
\end{figure}

Similarly, it can be easily seen that diagrams of the form
Fig.~\ref{fig:diagrams}c vanish if at most two of the three Higgs
fields are attached to the fermion loop. The only remaining
case is the one shown in Fig.~\ref{fig:dangerous}, where
all three Higgs fields are attached to the fermion loop. This case is
now 
discussed in more detail. 
Due to the structure $\qsl_i+m_i$ of the
numerators of the fermion propagators, the $\gamma$-chain $B$ in the
diagram of Fig.~\ref{fig:dangerous} is a sum of terms involving
between zero and five factors of 
$\qsl_i$, while the terms of the $\gamma$-chain $A$ involve either
zero or one such factor. 

After carrying out the fermion-loop integral, $B$ can contain up
to four covariants $\qsl_i\in\{\psl_{a,b,c},\ksl\}$, where $k$ is the
second loop momentum. Hence in those terms of $B$ that contain five
factors of $\qsl_i$, actually two of the $\qsl_i$ must be equal, and
the product of these five $\qsl_i$ can be reduced to products of only
three different $\qsl_i$. Hence such terms of $B$ contribute zero to
the diagrams.

Those terms of $B$ that contain four factors of $\qsl_i$ can
contribute only together with terms of $A$ containing no such
factor, due to property 1 above. In such a situation, the
$k$-dependence of the second loop integral has the following form,
\begin{align}
\int \frac{d^Dk}{(2\pi)^D}\,
\frac{\ksl\,f(k^2,kp_a,kp_b,kp_c)}
{[(k+p_a+p_b+p_c)^2-m_q^2][k^2-m_{\tilde{q}}^2]},
\label{secloop}
\end{align}
where the scalar function $f$ results from the fermion-loop
integral. As a result of the loop integral (\ref{secloop}), the factor
$\ksl$ within $B$ is replaced by a linear combination of
$\psl_{a,b,c}$. In this way,
each term within $B$ can be simplified to a term
involving at most two $\qsl_i$-factors. 

Taken together, there are no terms in the $\gamma$-chain $B$ that can
possibly give a contribution of diagrams of the form Fig.~\ref{fig:diagrams}c. 
Therefore, all diagrams in Fig.~\ref{fig:diagrams} vanish, and thus we obtain
\begin{align}
\Delta^{(2),\rm DRED}&=0 \, ,
\end{align}
and, with the help of (\ref{DeltaDRED}) and (\ref{phi4ctId2}), 
we find the  relation
\begin{align}
\delta^{(2)}V^{\rm quartic}_{\rm non-susy}
&=0 \, ,
\end{align}
which excludes the presence of non-symmetric counterterms.

\section{Conclusions}

We have considered the Yukawa-enhanced two-loop contributions to
Higgs-boson masses in the MSSM, i.e.\ contributions of
${\cal O}(\alpha_{t,b}\alpha_s)$, ${\cal O}(\alpha_{t,b}^2)$, 
${\cal O}(\alpha_t\alpha_b)$. In the literature these contributions
have been evaluated under the assumption that DRED is a supersymmetric
regularization up to the required order. In the present article we
have verified this assumption. 

The question of necessity for supersymmetry-restoring  counterterms
was studied 
utilizing the algebraic method of supersymmetric Slavnov-Taylor
identities.
A Slavnov-Taylor identity has been
found that unambiguously determines the potentially necessary
non-symmetric counterterms. 
It involves up to two-loop 5-point functions, but it
could be explicitly evaluated using the method of Ref.\
\cite{DREDPaper}. The identity turned out to be valid at the
regularized level, which means that
supersymmetry is preserved by DRED,
multiplicative renormalization is sufficient,
and therefore no extra supersymmetry-restoring
counterterms are necessary.

The restriction to the Yukawa-enhanced contributions and the
gauge-less limit is of crucial importance. 
If the gauge-less
limit is relaxed, more terms of the operator $[S(\Gamma^{(0)})]$ can
contribute and can lead to more
complicated Feynman diagrams. For example, there are diagrams with a
topology like in Fig.~\ref{fig:diagrams}c, but with a vector
boson instead of the virtual squark (and e.g.\ the
$\tilde{H}$--$q$--$\tilde{q}$-vertex replaced by a
$\tilde{H}$--$\tilde{H}$--$V$-vertex). The $\gamma$-strings $A$ and
$B$ in such diagrams have one 
$\gamma$-matrix more, and the arguments of section~\ref{sec:eval} 
cannot be applied.

It is, therefore, not straightforward to extrapolate 
the results of the
present paper to the full electroweak two-loop contributions and/or to
the three-loop level. The future experimental accuracy of Higgs-boson
mass measurements, however, requires to bring the full two-loop and
leading three-loop contributions under control. A dedicated study of
the supersymmetry-properties of DRED for these more complicated cases
will be an indispensable step in this program.

\begin{flushleft}

\end{flushleft}

\end{document}